\documentstyle[12pt]{article}

\def\be{\begin{eqnarray}}
\def\ee{\end{eqnarray}}
\def\nn{\nonumber}

\begin{document}

\phantom{.}\hfill ITEP-M-7/93 \\
\phantom{.}\hfill hep-th/9311142\\
\phantom{.}\hfill November 1993

\bigskip

\bigskip

\centerline{\Large{BOSONISATION OF COORDINATE RING OF $U_q(SL(N))$}}
\centerline{\Large{THE CASES OF $N=2$ and $N=3$}}

\bigskip

\centerline{A.Morozov}

\bigskip

\centerline{\it{117259 ITEP Moscow Russia}}

\bigskip

\bigskip

\bigskip

\centerline{ABSTRACT}

\bigskip

Non-abelian coordinate ring of $U_q(SL(N))$ (quantum deformation
of the algebra of functions) for $N=2,3$ is represented in terms
of conventional creation and annihilation operators.
This allows to construct explicitly representations of this
algebra, which were earlier described in somewhat more abstract
algebraic fashion.
Generalizations
to $N>3$ and Kac-Moody algebras are not discussed but look
straightforward.

\newpage

\section{Free-field representations of Lie algebras}

All algebras, relevant for applicatons in theoretical physics
usually possess the "free-field representations" which allow to
express all the generators of the algebra through creation and
annihilation operators, i.e. to embed the original algbera into
the universal envelopping of the (several copies of the)
Heisenberg algebra. This kind of representations for the 1-loop
(Kac-Moody, Virasoro, $W$- etc) algebras
\cite{FK,Wa,GMMOS} is the basis of the modern conformal field
theory. Their analogues for the ordinary (0-loop) Lie algebras
is also well known and widely used.

For the simplest case of $SL(2)$ this is the familiar representation:
\be
J^- = \frac{d}{dx}, \nn \\
J^0 = x\frac{d}{dx} - j, \nn \\
J^+ = x^2\frac{d}{dx} - 2jx,
\label{sl(2)}
\ee
and it can be considered as the "zero-mode" part of the Wakimoto
representation \cite{Wa,GMMOS} for $\hat{SL_k(2)}$,
\be
J^-(z) = W(z), \nn \\
J^0(z) = \chi W(z) -  q \partial\phi(z), \nn \\
J^+(z) = \chi^2W(z) - 2q \chi\partial\phi(z) - k\partial\chi(z), \nn \\
q = \sqrt{k+2}, \ \ \ W(z) = \frac{\delta}{\delta\chi(z)},
\label{Wakimoto}
\ee
which for $k=1$ turns into the standard Frenkel-Kac representation
\cite{FK} for $\hat{SL_1(2)}$ \cite{GMM},
\be
J^\pm(z) = {\cal C}_\pm e^{\pm\Phi(z)}, \nn\\
J^0(z) =   \partial\Phi(z).
\label{FK}
\ee

The free-field representations should and can be generalized to
include quantum groups. The analogue of {\ref{FK}} for $k=1$
and $q\neq 1$ is known as the Frenkel-Jing representation \cite{FJ}
(see also \cite{MV}),
generalizations of (\ref{Wakimoto}) for any $k$, though easy to derive,
are explicitly available only in the non-transparent terms of
three scalar fields
(instead of one scalar and $\beta,\gamma$ system),
see \cite{Boug} for a brief review.
The analogue of (\ref{sl(2)}) can be easily obtained in terms of the
finite-difference operators \cite{FV}, e.g.
\be
J_q^- = D^+, \nn \\
q^{\pm J_q^0} = q^{\mp j}M^\pm, \nn \\
J_q^+ = \frac{2q^{j+1}}{q+1} z^{2j+2}D^+z^{-2j} M^-.
\label{SL(2)_q}
\ee
Here
\be
D^\pm f(x) = \frac{f(x) - f(q^{\pm 1} x)}{(1- q^{\pm 1})x}, \nn \\
M^\pm f(x) = f(q^{\pm 1} x), \nn \\
M^\pm = I + (q^{\pm 1} - 1)xD^\pm.
\label{fideo}
\ee

All these formulas (and their analogues for any $N$)
can be easily deduced from the commutation relations for
the generators $T_\alpha$ of any Lie algebra by the following
procedure. Introduce
\be
{\cal F}({\bf x}) = \langle {\bf j} |
\prod_{\bf\alpha > {\bf 0}}\hat e_q\left(x_{\bf\alpha}
T_{-\bf\alpha}\right) = \langle {\bf j} |
\hat e_q\left(\sum_{\alpha > {\bf 0}}
f_q^\alpha({\bf x}) T_{-\alpha}\right),
\label{F-fun}
\ee
where $\alpha$ are somehow ordered labels of all the generators,
and the bra-vacuum $\langle {\bf j} |$ is annihilated by all the
``positive'' generators $T_\alpha$, $\alpha > 0$ and is the
eigenvector of all the ``Cartanian'' (mutualy commuting) ones,
$T_\alpha$, $\alpha
\in \{{\bf 0}\}$, eigenvalues being defined by the set ${\bf j}$.
For non-quantum groups $q=1$ and the $q$-exponent\footnote{
Let us remind that the $q$-exponent
$e_q(x) = 1/E_q(-x)$ is characterized by the
following set of properties:

1. $D^+ \hat e_q(x) = \hat e_q(x)$;
$\hat e_q(x) \equiv  e_q\left((1-q)x\right)$ and
$\lim_{q \rightarrow 1} \hat e_q(x) = e^x$;

2. $e_q(x) = \sum_{k \geq 0} \frac{x^k}{(q,q)_k}$,
$E_q(x) = \sum_{k \geq 0} \frac{q^{k(k-1)/2} x^k}{(q,q)_k}$, \\
where $(a,q)_k \equiv \prod_{i=0}^{k-1} (1-aq^i) =
\frac{(a,q)_\infty}{(aq^k,q)_\infty}$;

3. $e_q(x) = \frac{1}{(x,q)_{\infty}}$, thus
$E_q(x) = (-x,q)_\infty$  and
$\theta_{00}(x) \equiv \sum_{k = -\infty}^{\infty}q^{k^2/2}x^k =
(q,q)_\infty E_q(q^{1/2}x)E_q(q^{1/2}x^{-1})$;

4. $E_q(x) E_q(y) = E_q(x+y)$ and
$e_q(y) e_q(x) = e_q (x+y)$, provided $xy = qyx$;

5.$E_q(y) E_q(x) = E_q(x+y+yx)$ and
$e_q(x) e_q(y) =  e_q (x+y-yx)$, provided $xy = qyx$.

The first three properties explain the relevance of the $q$-special
functions as solutions to finite difference equations (i.e. to
various periodicity constraints), while the last two are crucial
for occurence of the same functions in the study of non-commutative
algebras and problems of quantum mechanics and quantum field theory.
}
$\hat e_q$ in (\ref{F-fun})
is substituted by the ordinary exponential function. Functions
$f^\alpha_1({\bf x})$ are polinomials in ${\bf x}$-variables, degree
of the polinomial being equal to the number of items in decomposition
of ${\bf\alpha}$ in the sum of the simple roots.
Representation of original algebra is now defined from the relation
\be
J_{\bf\alpha} {\cal F}({\bf x}) = {\cal F}({\bf x})T_{\bf\alpha}.
\label{gen-rep}
\ee
where $T_{\bf\alpha}$ at the r.h.s. is ``carried'' through the exponential
operator to act on the vacuum, and terms arising from commutation
of operators can be imitated by taking $x_{\bf \beta}$ derivatines.
Then
\be
J_{\bf\alpha}J_{\bf\beta} {\cal F}({\bf x}) =
J_{\bf\alpha}{\cal F}({\bf x})T_{\bf\beta} =
{\cal F}({\bf x})T_{\bf\alpha}T_{\bf\beta}.
\ee
In this way it is easy to derive not only (\ref{sl(2)}-\ref{SL(2)_q}),
but also all the other formulas from ref.\cite{GMMOS} and further
papers on free-field representations.

\section{Coordinate ring of the quantum group}

The purpose of this letter is to discuss the free-field representation
of the somewhat new object: the coordinate ring of the quantum group,
which is an essential piece of the theory but has a trivial classical
limit as $q=1$, where it becomes just a free abelian algebra.
For $q\neq 1$ this algebra is no longer abelian and provides a solution
to the basic equation \cite{FRT} ${\cal R} (T\otimes T) =
(T\otimes T){\cal R}$, where ${\cal R}$ is the R-matrix, i.e. solution
of the Yang-Baxter equation.

In the case of $U_q(SL(2))$
$T = \left(\begin{array}{cc} a & b \\ c & d \end{array}\right)$,
where the elements  $a, b, c, d$ (the $q\neq 1$ analogues
of matrix elements)
are no longer $c$-numbers, but operators with the following commutation
relations:
\be
ab = qba, \nn \\
ac = qca, \nn \\
ad - da = (q - q^{-1})bc, \nn \\
bc = cb, \nn \\
bd = qdb, \nn \\
cd = qdc.
\label{comrel}
\ee
This algebra can be also considered as that of the linear
automorphisms of the
``quantum phase plane'' \cite{Manin}, parametrized by the non-commuting
``coordinates''
\be
u_1u_2 = qu_2u_1
\label{quapla}
\ee
(they can be considered as exponentials of the coordinate
and momentum operators,
$u_1 = e^Q, \ u_2 = e^P, \ q = e^{ih}, \ P = -ih\frac{d}{dQ}$).

Commutation relations for the entries of $T$-matrices, associated
with $U_q(SL(N))$ can be easily described in terms of those
for $U_q(SL(2))$. If $T = \left(A_{ij}\right)$, $i,j = 1\ldots N$,
then for any fixed $i<k, \ j<l$ the $2\times 2$ matrix
$\left(\begin{array}{cc} A_{ij} & A_{il} \\ A_{kj} & A_{kl}
\end{array} \right)$
has exactly the same properties as
$\left(\begin{array}{cc} a & b \\ c & d \end{array} \right)$
in (\ref{comrel}) (i.e.
$A_{ij} A_{il} = q A_{il} A_{ij}$,
$A_{ij} A_{kl} - A_{kl} A_{ij} = (q - q^{-1}) A_{il} A_{jk}$ etc).

Below we describe representation of (\ref{comrel}) and its analogue
for $U_q(SL(3))$ in terms of annihilation and creation operators.
The case of arbitrary $N$ and Kac-Moody algebras will be discussed
elsewhere, as well as their relation to the more sophisticated
representations of the quantum group themselves (these are not
the same, of course: compare (\ref{comrel}) and (\ref{SL(2)_q})).
Instead we discuss briefly representations of the coordinate ring,
which were earlier found in pure algebraic terms in \cite{GLS,Kar}
and can be now constructed explicitly in terms of functions of
commuting variables.

\section{Oscillator representation of the basic algebras}

Before addressing the question of bosonization of the algebra
(\ref{comrel}), we consider the even simpler ``quantum hyperplane''
algebras (compare with (\ref{quapla}). These will provide us with
the building blocks for further constructions.
We shall need two such algebras: the ``chain'' (or ``quantum phase
space'') one,
\be
w_i w_j = q w_j w_i\ \ {\rm for} \ j = i+1, \nn \\
w_i w_j = w_j w_i \ \ {\rm for} \ |j-i| > 1,
\label{chain}
\ee
and the ``huperplane'' one
\be
u_i u_j = q u_j u_i \ \ {\rm for\ any} \ j>i.
\label{quhyp}
\ee
Free-field representation (bosonization) expresses these generators
through those of Heisenberg algebra,
\be
[\alpha_i, \alpha_j^\dagger] = \delta_{ij}\log q.
\label{Heis}
\ee
Such representation for (\ref{chain}) is straightforward:
\be
w_i = e^{\alpha^\dagger_{i-1} + \alpha_i}.
\label{bos-chain}
\ee
That for (\ref{quhyp}) can be obtained from (\ref{bos-chain}):
\be
u_i = \ : \prod_{k \leq i} w_k : \ = \ :
\exp\left(\sum_{k < i} \alpha^\dagger_k +
\sum_{k \leq i} \alpha_k\right):
\label{bos_quhyp}
\ee

All the operators involved are exponentials of linear combinations
of creation and annihilation operators and normal ordering can be
defined by just requesting that whenever the Wick theorem is applied
for evaluation of correlation functions and/or commutation
relations, no contractions are included of operators standing
under the normal ordering signs. For any such operators
\be
{\cal O} = \ :\exp\left(\sum_k A_k\alpha_k + \sum_k
B_k \alpha^\dagger_k\right):
\label{O-ops}
\ee
we have:
\be
{\cal O}_1\cdot {\cal O}_2 =
\sqrt{\epsilon_{12}}:{\cal O}_1{\cal O}_2:\ =
\epsilon_{12}{\cal O}_2\cdot{\cal O}_1
\label{O-comrel}
\ee
where the $c$-number $\epsilon_{12} =
q^{\sum_k \left(A_k^{(1)}B_k^{(2)} - B_k^{(1)}A_k^{(2)}\right)}$.
Since all the operators below will be of the form
(\ref{O-ops}), in what follows we use (\ref{O-comrel})
without special reference.

\section{The case of $U_q(SL(2)$}.

We can now describe bosonization formulas for the algebra
(\ref{comrel}). For this purpose we need two mutualy commuting
copies of algebra (\ref{quhyp}), their generators will be denoted
by $\{u_i\}$ and $\{v_i\}$, $u_iv_j = v_ju_i$.
Then (\ref{comrel}) is immediately satisfied, if
\be
\left(\begin{array}{cc} a & b \\ c & d \end{array}\right) =
\left(\begin{array}{cc} u_1v_1 & u_1v_2 \\ u_2v_1 & u_2v_2
\end{array}\right).
\label{bos-SL(2)_q}
\ee
The only relation which is a little non-trivial is
$ad - da = (q - q^{-1})bc$. It is at this place that
(\ref{O-comrel}) plays the crucial role. Indeed,
\be
ad = \ : u_1v_1:\ \cdot \ : u_2v_2: \ = q:u_1v_1u_2v_2:, \nn \\
da = \ : u_2v_2:\ \cdot \ : u_1v_1: \ = q^{-1}:u_1v_1u_2v_2:, \nn \\
bc = \ : u_2v_1:\ \cdot \ : u_1v_2: \ = :u_1v_1u_2v_2: = cb.
\ee

Representation (\ref{bos-SL(2)_q}) has an obvious generalization
for $U_q(SL(N))$ with any $N$: it is enough to take
\be
T = \left( A_{ij} \right) = \left( u_i v_j \right).
\label{bos-SL(N)_q}
\ee

However, both (\ref{bos-SL(2)_q}) and (\ref{bos-SL(N)_q}) are
{\it non-generic} representations: they are
actually degenerate. This is clear,
because the $c$-number $D = \det_q T = ad - qbc = da - q^{-1}bc$
in the case of (\ref{bos-SL(2)_q}) is identically vanishing:
$D=0$. In the case of (\ref{bos-SL(N)_q}) the situation is even
worse: not only the full determinant of the matrix T, but also
all its minors are identiically vanishing. We now proceed to
description of generic, non-degenerate representations.

For the case of $U_q(SL(2))$ it is very simple to introduce the
necessary correction: instead of (\ref{bos-SL(2)_q}) one can
take
\be
T =
\left(\begin{array}{cc} a & b \\ c & d \end{array}\right) =
\left(\begin{array}{cc} u_1v_1 & u_1v_2 \\ u_2v_1 & u_2v_2 +
D_2\frac{1}{u_1v_1}
\end{array}\right).
\label{bosmod-SL(2)_q}
\ee
Here $D_2 = \det_q T$ is a $c$-number,
commuting with all the operators $u$ and
$v$. Since $u$'s and $v$'s are of the form (\ref{O-ops}) there are no
problems with the definition of their negative powers. Using
(\ref{quhyp}) along with its obvious corollary,
\be
u_i \frac{1}{u_j} = q\frac{1}{u_j} u_i, \ {\rm for} \ i>j,
\ee
it is easy to check that all the relations (\ref{comrel}) are still
true  for representation (\ref{bosmod-SL(2)_q}).

\section{The case of $U_q(SL(3))$}.

The non-degenerate representation now looks as follows:
\be
T = \left(\begin{array}{ccc} a & b & e \\ c & d & f \\ g & h & k
\end{array}\right) =
\left( \begin{array}{ccc} u_1v_1 & u_1v_2 & u_1v_3 \\
u_2v_1 & u_2v_2\cdot W &  u_2v_3 \cdot W \\
u_3v_1 & u_3v_2 \cdot W & u_3v_3 \cdot W + D_3:\frac{u_4v_4}{u_3v_3}:
\end{array}\right); \nn \\
W = 1 + \frac{1}{q}:\frac{u_3v_3}{u_1v_1u_2v_2u_4v_4}:
\label{bosm-SL(3)_q}
\ee
or
\be
T = \left( \begin{array}{ccc} u_1v_1 & u_1v_2 & u_1v_3 \\
u_2v_1 & u_2v_2 +  :\frac{u_3v_3}{u_1v_1u_4v_4}: &  u_2v_3 +
                   :\frac{u_3v_3^2}{u_1v_1v_2u_4v_4}: \\
u_3v_1 & u_3v_2 +  :\frac{u_3^2v_3}{u_1v_1u_2u_4v_4}: &
         u_3v_3 +  :\frac{u_3^2v_3^2}{u_1v_1u_2v_2u_4v_4}: +
                 D_3:\frac{u_4v_4}{u_3v_3}:
\end{array}\right)
\label{bosmod-SL(3)_q}
\ee

The $q$-determinant of the entire matrix $T$ is equal to
$\det_q T = D_3$ and commutes with all the entries of $T$.
The extra terms, appering in expressions for $d, f, h, k$
are important to make the $2\times 2$ minors
of $T$ non-degenerate, for example:
\be
\Delta_{33} \equiv ad - qbc = \ :\frac{u_3v_3}{u_4v_4}:
\label{Deltas}
\ee
It is clear that $\Delta_{33}$ commutes with all
the other
contributions to the elements $a, b, c, d$ of the $U_q(SL(2))$
subalgebra. Representation of this subalgebra reduces to
(\ref{bosmod-SL(2)_q}) if $\Delta_{33}$ is identified with $D_2$.
However, in $U_q(SL(3))$ $\Delta_{33}$ is no longer a $c$-number,
but it still has simple (multiplicative-like) commutation
relation with $k$:
\be
\Delta_{33} k - q^2k\Delta_{33} = (1-q^2)D_3
\label{Del-k}
\ee
Such simple relations arise as a rule for $q$-commutators of the
minors for all the $U_q(SL(N))$
algebras \cite{Kar}.

The check that all the commutation relations of $U_q(SL(3)$ are
satisfied for representation (\ref{bosmod-SL(3)_q}) is a somewhat
tedious but straightforward exercise. Again one should repeatedly
make use of the relation (\ref{O-comrel}). As an example, let us
check that $fk = qkf$. This is true since
\be
fk = \left(u_2v_3 +  :\frac{u_3v_3^2}{u_1v_1v_2u_4v_4}:\right)\cdot
  \left(u_3v_3 +  :\frac{u_3^2v_3^2}{u_1v_1u_2v_2u_4v_4}: +
                 D_3:\frac{u_4v_4}{u_3v_3}:\right) = \nn\\
= q^{1/2}:u_2u_3v_3^2:\ + \
(q^{3/2} + q^{-1/2}):\frac{u_3^2v_3^3}{u_1v_1v_2u_4v_4}:\ + \nn \\ +\
q^{1/2}:\frac{u_3^3v_3^4}{u_1^2v_1^2u_2v_2^2u_4^2v_4^2}: + \
q^{1/2}D_3\left(:\frac{u_2u_4v_4}{u_3}: +
:\frac{v_3}{u_1v_1v_2}:\right)
\ee
while
\be
kf = q^{-1/2}:u_2u_3v_3^2:\ + \
(q^{-3/2} + q^{1/2}):\frac{u_3^2v_3^3}{u_1v_1v_2u_4v_4}:\ + \nn \\ +\
q^{-1/2}:\frac{u_3^3v_3^4}{u_1^2v_1^2u_2v_2^2u_4^2v_4^2}: + \
q^{-1/2}D_3\left(:\frac{u_2u_4v_4}{u_3}: +
:\frac{v_3}{u_1v_1v_2}:\right)
\ee

\section{On representation theory of coordinate rings}

According to \cite{GLS,Kar} representations of the $U_q(SL(N))$
can be described in terms of vectors, obtained by the action of
certain ``creation operators''\footnote{Not to be mixed with
the Heisenberg operators $\alpha^\dagger,\ \beta^\dagger$.}
on the ``vacuum''. The ``vacuum'' is
defined as the common eigenvector of the maximum set of commuting
generators of coordinate ring, while ``creation operators'' are
certain minors of the matrix $T$ \cite{Kar}. Realization of $T$ in terms
of the generators of Heisenberg algebra allows to construct all
these objects explicitly. We present here  only the example of
$U_q(SL(2))$.

The maximum set of commuting generators for generic $q$ consists of
$b$ and $c$. Thus ``vacuum'' state is defined to satisfy
\be
b |vac\rangle = \mu |vac\rangle, \ \ c|vac\rangle = \nu |vac\rangle.
\label{vac}
\ee
In order to obtain a highest weight representation one also requires
that
\be
d |vac\rangle = 0,
\label{hw}
\ee
and the entire representation is formed by the vectors
\be
a^n |vac\rangle.
\label{elrep}
\ee
For special values of $q = e^{2\pi i/p}$ with integer $p$ the set of
commuting operators is actually bigger: $d^p$ commutes with both $b$
and $c$ and instead of (\ref{hw}) ``vacuum'' can be defined to be
also an eigenstate of $d^p$:
\be
d^p|vac\rangle = \delta|vac\rangle,\ \ {\rm for} \ q^p = 1,\ p\in Z.
\label{cirep}
\ee

Using (\ref{bosmod-SL(2)_q}) we can now represent $a, b, c, d$
through a pair of Heisenberg creation and annihilation operators:
\be
a = e^{\alpha + \beta}, \ \
b = e^{\alpha^\dagger + \beta}, \ \
c = e^{\beta^\dagger + \alpha}, \nn \\
d = e^{\alpha^\dagger + \beta^\dagger} + D_2 e^{-\alpha - \beta}.
\ee
It is now possible to represent Heisenberg generators in terms of
differential operators:
\be
\alpha^\dagger = x\log q, \ \ \alpha = \frac{d}{dx}, \nn \\
\beta^\dagger = y\log q, \ \ \beta = \frac{d}{dy}.
\ee
Then $a, b , c, d$ acquire the form of finite-difference operators:
\be
a = m^+_x m^+_y,\ \ b = q^ym^+_x,\ \ c = q^xm^+_y, \ \
d = q^{x+y} + D_2m^-_xm^-_y, \nn \\
m^\pm_x \equiv e^{d/dx}, \ \ m^\pm_x f(x) = f(x\pm 1).
\label{bosope}
\ee
Therefore as long as we deal with coordinate ring only and
not with the Heisenberg algebra itself $x$ and $y$ can be considered
as variables on the integer lattice: $x, y \in Z$.
Solution to eqs.(\ref{vac}) is now given by
\be
|vac\rangle \sim q^{-xy}\mu^x\nu^y \equiv |\mu, \nu \rangle.
\label{munuvac}
\ee
Conditions (\ref{hw}) or (\ref{cirep}) can be now considered as
definitions of $D_2$ in (\ref{bosope}) in terms of $\mu,\ \nu$ and
$\delta$: (\ref{hw}) implies that
\be
D_2 = -q\mu\nu,
\ee
while (\ref{cirep}) means that
\be
\delta = \prod_{k=1}^p \left(1 + \frac{D_2}{q^{2k+1}\mu\nu}\right).
\ee
Representation itself consists of the states
\be
|n\gg \equiv a^n|vac> \sim q^{-(x+n)(y+n)}\mu^{x+n}\nu^{y+n} =
q^{-n^2}\mu^n\nu^n | q^{-n}\mu,q^{-n}\nu\rangle.
\label{repel}
\ee
For $q = e^{2\pi i/p}$ there are at most $p$ linearly independent
states in this representation. Moreover, for even $p$ there are actualy
irreducible representations of the size $p/2$. This is clear from the
fact that all the states (\ref{repel}) are eigenstates of $b$ and $c$,
and action of operators $a$ and $d$ in (\ref{bosope})
does not change parity
of the integer-valued combination $x+y$. Thus representation
(\ref{repel}) can be defined on the sublattice $x,y \in Z$,
$x+y \in 2Z$ and $|-\mu,-\nu\rangle$ can be identified with
$|\mu,\nu\rangle$.

\section{Acknowledgements}

I am indebted to A.Bougourzi, V.Karimipour and L.Vinet
for explanations and discussions
which stimulated my interest to this subject.

\end{document}